\documentclass{moriond}

\usepackage{amssymb}

\def\be{\begin{equation}}
\def\ee{\end{equation}}
\def\bea{\begin{eqnarray}}
\def\eea{\end{eqnarray}}

\newcommand{\Photo}{}

\begin{document}
\vspace*{4cm}
\title{A Natural extra-dimensional origin for the LHCb anomalies}

\author{E.~Meg\'\i as$^{1,2}$, G.~Panico$^3$~\footnote{Speaker.}, O. Pujol\`as$^3$ and M.~Quir\'os$^3$}

\address{\vspace{1.em}{}$^1$Departamento de F\'\i sica Te\'orica, Universidad del Pa\'\i s Vasco UPV/EHU,\\ Apartado 644, 48080 Bilbao, Spain\\
{}$^2$Max-Planck-Institut f\"ur Physik (Werner-Heisenberg-Institut),\\ F\"ohringer Ring 6, D-80805, Munich, Germany\\
{}$^3$IFAE and BIST, Universitat Aut\`onoma de Barcelona, 08193~Bellaterra,~Barcelona, Spain \vspace{.5em}}

\maketitle\abstracts{
We study the possibility of explaining the recently found anomalies in $B$-meson decays within
scenarios with a composite Higgs boson. This class of models provides a natural way to fit the experimental results,
interpreting the anomalies as the result of the exchange of heavy vector resonances with electroweak quantum numbers.
The anomalies are tightly related to deviations in $\Delta F = 2$ transitions and to deformations of the $Z$ and $W$ couplings,
whose size is of the order of the present experimental bounds. This leads to a very predictive scenario which could be soon
tested at collider experiments.
}

\section{Introduction}

Flavor observables can provide excellent probes of beyond the Standard Model (SM) physics. In particular rare $B$-meson
decays due to the $b \rightarrow s \ell^+ \ell^-$ transition, which are loop and CKM suppressed in the SM, can be
tested with good accuracy at b-factories and at the LHC, and constitute a privileged channel to test
the lepton flavor universality (LFU) hypothesis.

Recently the LHCb collaboration measured the ratio of the $B$ decays into a $K^{+,*}$
and a pair of muons or electrons~\cite{Aaij:2014ora,RK*}, finding deviations of $\sim 2.5\,\sigma$ from
the SM predictions ($R_{K^{(*)}} = 1$)
\begin{eqnarray}
&\displaystyle R_K = \frac{{\rm BR}(B^+ \rightarrow K^+ \mu^+ \mu^-)}{{\rm BR}(B^+ \rightarrow K^+ e^+ e^-)} = 0.745^{+0.090}_{-0.074} \pm 0.036
\hspace{2.5em}
1\; {\rm GeV}^2 < q^2 < 6\; {\rm GeV}^2\,,\\
&\displaystyle R_{K^*} = \frac{{\rm BR}(B^+ \rightarrow K^* \mu^+ \mu^-)}{{\rm BR}(B^+ \rightarrow K^* e^+ e^-)} =
\left\{
\begin{array}{l@{\hspace{2em}}l}
0.660^{+0.110}_{-0.070}\pm 0.024 & (2 m_\mu)^2 < q^2 < 1.1\; {\rm GeV}^2\\
\rule{0pt}{1.25em}0.685^{+0.113}_{-0.069}\pm 0.047 & 1.1\; {\rm GeV}^2 < q^2 < 6\; {\rm GeV}^2
\end{array}
\right.\,.
\end{eqnarray}
These results are quite intriguing since they are obtained in `clean' channels with low theoretical uncertainties. They seem to point
towards a sizable violation of LFU.

Additional deviations from the SM predictions have also been found in related observables, namely the semi-leptonic branching ratios of $B \rightarrow K^{(*)} \mu^+\mu^-$ and $B_s \rightarrow \phi \mu^+\mu^-$ and in the angular distributions
of the decay $B \rightarrow K^*\mu^+ \mu^-$ (in particular the $P_5'$
observable).~\cite{Aaij:2014pli}

A departure from LFU in $b \rightarrow s \ell^+ \ell^-$ decays may be due to non-universal
new-physics contributions to the effective operators
\begin{equation}
{\cal O}_9^{(\prime)\ell} = (\overline s_{L,R} \gamma_\mu b_{L,R}) (\bar\ell \gamma^\mu \ell)\,,
\qquad\quad
{\cal O}_{10}^{(\prime)\ell} = (\overline s_{L,R} \gamma_\mu b_{L,R}) (\bar\ell \gamma^\mu\gamma_5 \ell)\,.
\end{equation}
The fit to the present data prefers a negative shift in $C_9^\mu$, possibly correlated to a positive contribution
to $C_{10}^\mu$.~\cite{Capdevila:2017bsm,DAmico:2017mtc} This pattern of deviations can be explained if new physics is present
that couples dominantly to the muon field.

Several theoretical analyses proposed interpretations of the anomalies within a BSM perspective.
The most obvious possibilities are extensions of the SM involving new massive $Z'$ bosons or leptoquarks.
A shortcoming of many of these constructions is the fact that the BSM dynamics has no fundamental reason for being
present, other than explaining the $B$ anomalies. In the following we use a different approach: we do not add
ad-hoc new states, but instead we try to connect the LHCb anomalies to some BSM dynamics whose main motivation
is addressing the EW Hierarchy Problem. A natural way to do this, as we will discuss in the following sections, is to
focus on BSM scenarios with a composite Higgs and a new strongly-coupled dynamics.~\cite{Niehoff:2015bfa,Megias:2016bde,Megias:2016jcw}

\section{The $B$-meson anomalies in composite Higgs scenarios}\label{sec:power-counting}

In this section we provide a power-counting analysis of the $B$-meson anomalies in the context of
composite Higgs theories. In order to keep the discussion as general as possible, we will
not specify whether the Higgs is a generic composite ``mesonic'' state
(as in Randall-Sundrum (RS) scenarios) or is a (pseudo-)Goldstone boson. Although the estimates can vary by order one factors,
the main qualitative features remain the same in the two scenarios.

In theories with new strongly-coupled dynamics two natural candidates can give rise to $\Delta F = 1$ effective operators
involving the $b$ and $s$ quarks: the exchange of heavy vector resonances with electroweak (EW) quantum numbers
(analogous to $Z'$ states), and the presence of flavor-changing interactions of the SM $Z$ boson~\cite{Niehoff:2015bfa}.

\begin{figure}
\centering
\includegraphics[height=.16\textwidth]{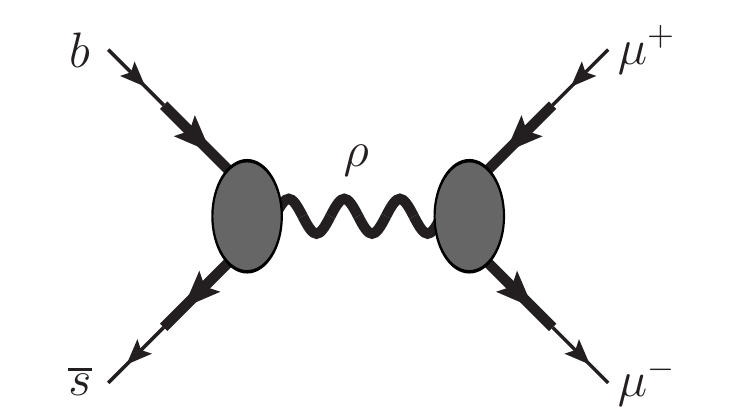}
\hfill
\includegraphics[height=.16\textwidth]{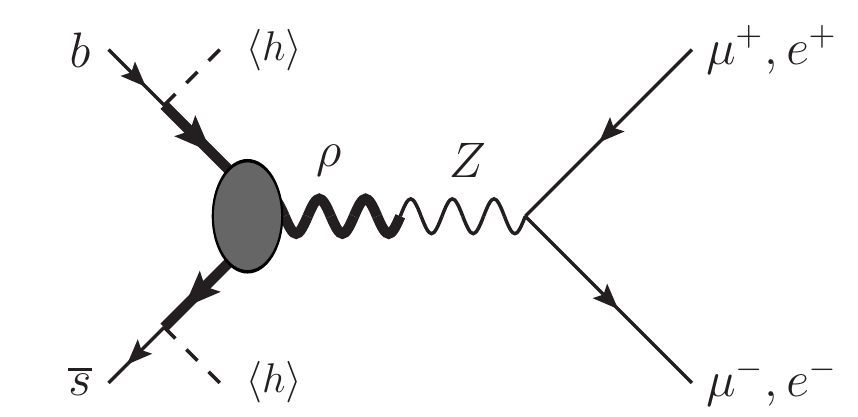}
\hfill
\includegraphics[height=.16\textwidth]{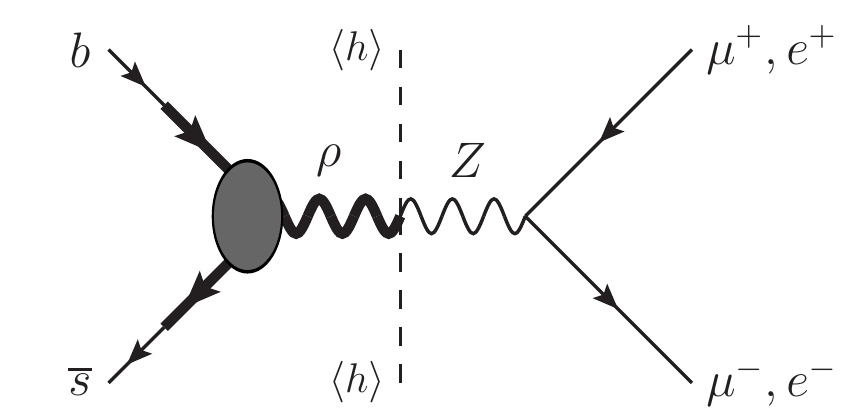}
\vspace{-.3cm}
\caption{Schematic structure of the diagrams giving rise to contributions to the ${\cal O}_{9, 10}$
effective operators through the exchange of a heavy vector resonance (left panel) or through
a flavor changing $Z$-boson coupling (middle and right panel).
The thick lines denote composite resonances coming from the strongly coupled dynamics.}\label{fig:c9}
\end{figure}

We start by discussing the former effect, whose origin can be qualitatively understood from the diagram
in the left panel of fig.~\ref{fig:c9}. In principle, contributions to all ${\cal O}^{(\prime)}_{9,10}$ operators
can be present. The size of each contribution is determined by the amount of compositeness (i.e.~the size of the mixing
with the composite resonances) of the $L$ and $R$-handed chiralities of the $b$ quark and of the muon.
In general one expects the $b_L$ field to have a sizable amount of compositeness, since
it forms a doublet with the $t_L$ field. The large top Yukawa requires both top chiralities to be strongly mixed
with the composite dynamics. The $b_R$ component, on the other
hand, has typically a small mixing with the composite states, since its compositeness is related to the
size of the bottom Yukawa. This pattern of compositeness implies that the largest new physics effects are
expected in the ${\cal O}_{9, 10}$ operators, while ${\cal O}'_{9, 10}$ are typically smaller.

Let us now focus on the lepton sector. Since the experimental data seem to point towards a violation of LFU,
we assume that the muon and the electron have different amount of compositeness. In particular, the safest option is to
assume that the electron is an almost elementary state with tiny compositeness, whereas the muon compositeness can
be sizable. In this scenarios one generates only contributions to ${\cal O}_{9,10}^\mu$ and not to ${\cal O}_{9,10}^e$.
The small size of the muon Yukawa, tells us that in natural scenarios only one muon chirality can have a large
compositeness. If the $\mu_R$ is a composite state one gets new-physics contributions that follow the pattern
${\cal O}_9^{\mu} = {\cal O}_{10}^\mu$. This possibility is strongly disfavored by the data~\cite{Capdevila:2017bsm,DAmico:2017mtc}.
The other option is to
assume a sizable compositeness for the $\mu_L$, leading to ${\cal O}_9^{\mu} = - {\cal O}_{10}^\mu$,
which can provide a very good fit to the experimental anomalies.~\footnote{The pattern of compositeness we described before
is almost mandatory in models with a Goldstone Higgs. For instance
it is directly realized in the quark sector in anarchic partial compositeness scenarios. In models with a ``mesonic'' Higgs,
instead, large mixings of all the chiralities with the composite dynamics are possible, although they need to be compensated
by `unnaturally' small values of the Yukawa couplings of the Higgs with the composite partners.}

The vector resonances contributions to the ${\cal O}_{9,10}^\mu$ operator coefficients can be estimated
as~\cite{Niehoff:2015bfa,Panico:2015jxa}
\begin{equation}\label{eq:C9}
\Delta C_9^\mu \simeq - \Delta C_{10}^\mu \sim -\frac{\sqrt{2} \pi}{G_F \alpha_{em}} s_{b_L}^2 s_{\mu_L}^2 \left(\frac{g_\rho}{m_\rho}\right)^2
\simeq - 0.4 \left(\frac{1\ {\rm TeV}}{m_\rho/g_\rho}\right)^2 \left(\frac{s_{b_L}}{0.3}\right)^2 \left(\frac{s_{\mu_L}}{0.3}\right)^2\,,
\end{equation}
where $s_{b_L}$ and $s_{\mu_L}$ parametrize the sine of the mixing angle between the $b_L$ and $\mu_L$ fields and the
composite partners, $m_\rho$ is the mass of the vector resonances, while $g_\rho$ is the size of the coupling characterizing
the strongly-coupled dynamics. To obtain the above estimate we assumed that the rotation matrix that diagonalizes the
down-type quark masses is approximately given by the CKM matrix, as happens in generic composite Higgs
scenarios.
One can see that the estimate in eq.~(\ref{eq:C9}) can easily reproduce the values needed to explain the anomalies in
$B$-meson decays~\cite{Capdevila:2017bsm}
\begin{equation}
\Delta C_9^\mu = - \Delta C_{10}^\mu = -0.61 \quad \textrm{best\ fit}\,, \qquad
\Delta C_9^\mu = - \Delta C_{10}^\mu \in [-0.87, -0.36]\quad \textrm{at}\ 2\,\sigma\;\textrm{C.L.}
\end{equation}
A good agreement with the fit requires vector resonances with a mass $m_\rho \sim \textit{few}\ {\rm TeV}$ and
a $b_L$ and $\mu_L$ compositeness $s_{b_L} \sim s_{\mu_L} \sim 0.3$.

A second set of new physics contribution to the ${\cal O}_{9,10}^{(\prime)}$ operators can come from
flavor changing currents mediated by the $Z$ boson.
The flavor changing effects can be induced after EW symmetry breaking (EWSB) by the mixing of the $Z$-boson
with heavy vector resonances and by the mixing of the SM fermions with composite partners with different quantum numbers.
Examples of diagrams giving rise to these effects are shown in the middle and right panel of fig.~\ref{fig:c9}.
Since the coupling with the leptons is due to the SM $Z$ current, these effects give rise to lepton-flavor universal
contributions. In particular the largest contributions are the ones to the coefficients of the  ${\cal O}_{10}^{e, \mu}$
operators, whose size can be estimated as~\cite{Niehoff:2015bfa,Panico:2015jxa}
\begin{equation}\label{eq:C10}
\Delta C_{10}^e = \Delta C_{10}^\mu \sim \frac{\sqrt{2} \pi}{G_F \alpha_{em}} s_{b_L}^2 \frac{g_\rho^2}{m_\rho^2}\,.
\end{equation}
Notice that contributions to $C'_{10}$ are proportional to the $b_R$ compositeness angle, so they are typically suppressed.
Moreover contributions to the ${\cal O}_{9}^{(\prime)}$ operators are accidentally small due to the smallness of the
vector coupling of the $Z$ to charged leptons, which is suppressed by a factor $1 - 4 \sin^2 \theta_{\textsc{w}} \simeq 0.08$
with respect to the axial coupling.

\begin{figure}
\centering
\includegraphics[height=.16\textwidth]{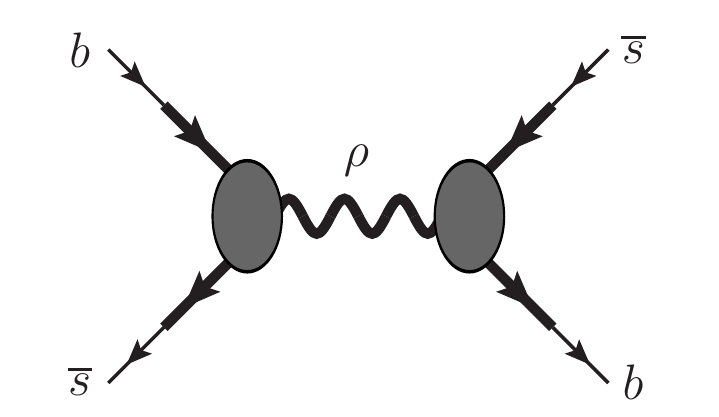}
\hspace{3em}
\includegraphics[height=.16\textwidth]{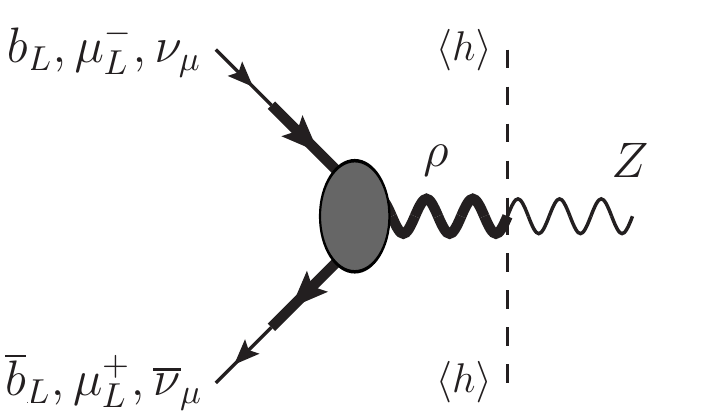}
\hspace{2em}
\includegraphics[height=.16\textwidth]{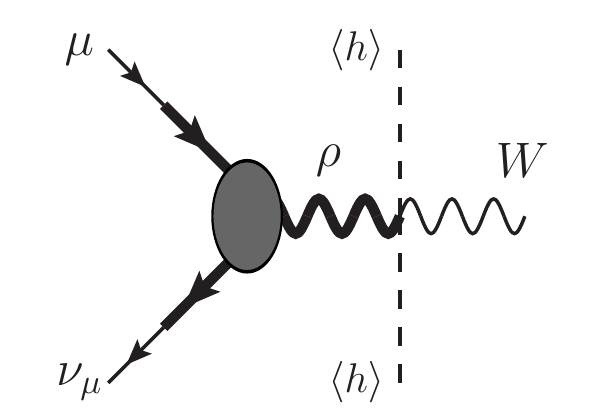}
\vspace{-.3cm}
\caption{Schematic structure of the diagrams giving rise to contributions to $\Delta F = 2$ transitions (left panel)
and to distortions of the $Z$ and $W$ couplings (middle and right panels).
Additional contributions to the distortions of the $Z$ and $W$ couplings can be generated by Higgs insertions in the mixing
of the quarks with the composite partners (see fig.~\ref{fig:c9}).}\label{fig:deltaF2}
\end{figure}

Since the contributions in eq.~(\ref{eq:C10}) are lepton-flavor universal, they do not modify the $R_K$ and $R_{K^*}$
observables, so they play a marginal role in fitting the $B$ anomalies and, for simplicity, we will not take them into account in the fit.
Notice moreover that these effects are directly related to the modifications of the $Zb_L\overline b_L$ coupling
(see fig.~\ref{fig:deltaF2}), whose size can be estimated as
\begin{equation}\label{eq:deltaZbb}
\frac{\delta g_{Zb_Lb_L}}{g_{Zb_Lb_L}^{\textsc{sm}}} \sim s_{b_L}^2 \frac{g_\rho^2}{m_\rho^2} \frac{v^2}{2}\,,
\end{equation}
where $v = \langle h \rangle \simeq 246\ {\rm GeV}$ is the Higgs vacuum expectation value. The current bounds on the deviations of the
$Zb_L\overline b_L$ coupling are of order
$|\delta g_{Zb_Lb_L}/g_{Zb_Lb_L}^{\textsc{sm}}| \lesssim 10^{-3}$.~\cite{Grojean:2013qca}~\footnote{To estimate the bound
we took into account the fact that deviations in the $Zb_R\overline b_R$ coupling are small due to the small $b_R$ compositeness.}

By using $\delta g_{Zb_Ls_L} \simeq V_{ts} \delta g_{Zb_Lb_L}$ we can translate the bound on the
$Zb_L\overline b_L$ deviations into an upper bound on the contributions to the ${\cal O}_{10}^{e,\mu}$ operators:
\begin{equation}
|\Delta C_{10}^{e, \mu}| \sim \frac{\sqrt{2}\pi}{G_F \alpha_{em}} \frac{1}{v^2} \frac{\delta g_{Zb_Lb_L}}{g^{\textsc{sm}}_{Zb_Lb_L}} \lesssim 1\,,
\end{equation}
which tells that these effects are not dangerously large.

It must be noticed that the $Z$ couplings to down-type quarks can be protected
by imposing a $P_{LR}$ symmetry~\cite{Agashe:2006at}. This symmetry reduces
the deviations in the $Zb_L\overline b_L$ coupling as well as the flavor-changing interaction $Z b_L s_L$,
hence it naturally suppresses the contributions to ${\cal O}_{10}^{e,\mu}$.

\medskip

The corrections to the ${\cal O}_{9,10}$ operators are also directly connected to additional new-physics effects,
whose size is strongly constrained experimentally. An unavoidable effect is the generation of contributions to $\Delta F = 2$
flavor-changing transitions involving down-type quarks. As shown in the left panel of fig~\ref{fig:deltaF2},
the exchange of vector resonances give rise to the effective operators
\begin{equation}
{\cal O}^{LL}_{\Delta F =2} \sim  s_{b_L}^4\!\! \left(\frac{g_\rho}{m_\rho}\right)^{\!2}\! (V^*_{3i} V_{3j})^2
(\bar d_{iL} \gamma^\mu d_{jL})^{2}
= \frac{1}{(10\ {\rm TeV})^2}\left(\frac{s_{b_L}}{0.3}\right)^{\!4}\! \left(\frac{1\ {\rm TeV}}{m_\rho/g_\rho}\right)^{\!2}\!
 (V^*_{3i} V_{3j})^2 (\bar d_{iL} \gamma^\mu d_{jL})^2\,.
\end{equation}
Notice that these operators can also be induced by colored vector resonances, which in RS scenarios
typically give rise to the largest contributions. The values of $m_\rho$ and $s_{b_L}$ required to explain the $B$-anomalies
give rise to contributions to $\Delta F = 2$ processes not far from the present bounds
$C_{\Delta F =2}^{LL} \lesssim 1/(5\ {\rm TeV})^2$.
Additional contributions can also be generated for the $\Delta F = 2$
operators with $LR$ and $RR$ chiralities. These are however not very large since the $b_R$ compositeness is
relatively small.

As we mentioned before, the presence of vector resonances and the sizable $b_L$ compositeness can give
rise to deviations in the $Zb_L\overline b_L$ couplings. Analogous effects are there for the muon. In the absence of a
custodial $P_{LR}$ protection we expect the $Z\mu_L\overline \mu_L$ coupling to acquire corrections
\begin{equation}\label{eq:deltaZmumu}
\frac{\delta g_{Z\mu_L\mu_L}}{g_{Z\mu_L\mu_L}^{\textsc{sm}}} \sim s_{\mu_L}^2 \frac{g_\rho^2}{m_\rho^2} \frac{v^2}{2}\,.
\end{equation}
The deviations in this coupling are constrained to be $|\delta g_{Z\mu_L\mu_L}/g_{Z\mu_L\mu_L}^{\textsc{sm}} | < 5\times 10^{-3}$.~\cite{Olive:2016xmw} Thus they can give a strong bound on the $\mu_L$ compositeness.
Notice that, since also the $\nu_\mu$, which belongs to the same $SU(2)_L$ multiplet as the $\mu_L$, has a sizable compositeness,
the couplings $Z\nu_\mu\overline \nu_\mu$ and $W\!\mu\nu_\mu$ acquire corrections of the order
\begin{equation}
\frac{\delta g_{W\!\mu\nu_\mu}}{g_{W\!\mu\nu_\mu}^{\textsc{sm}}} \sim
\frac{\delta g_{Z\nu_\mu\nu_\mu}}{g_{Z\nu_\mu\nu_\mu}^{\textsc{sm}}} \sim
s_{\mu_L}^2 \frac{g_\rho^2}{m_\rho^2} \frac{v^2}{2}\,.
\end{equation}
These couplings can be bounded from the measurement of the Fermi constant in muon decays and from the LEP measurement
of the invisible $Z$ width.~\cite{Niehoff:2015bfa}
Both constraints give bounds of the order of $\textit{few}\times 10^{-3}$.
It is interesting to notice that, if the $Z\mu_L\overline \mu_L$ couplings are protected by the
custodial $P_{LR}$ symmetry, the couplings involving the neutrinos can not have such protection at the same time.
Thus the bound on the $\mu_L$ compositeness is unavoidable in these scenarios.

Comparing the estimate of the contributions to $C_{9, 10}^\mu$ in eq.~(\ref{eq:C9}) with the size of the deviations
in the $Zb_L\overline b_L$ and $Z\mu_L\overline \mu_L$ couplings in eqs.~(\ref{eq:deltaZbb}) and (\ref{eq:deltaZmumu})
we find
\begin{equation}
\Delta C_9^\mu \simeq - \Delta C_{10}^\mu \sim -0.4 \left(\frac{m_\rho/g_\rho}{1\ {\rm TeV}}\right)^2 \left(\frac{\delta g_{Zb_Lb_L}/g_{Zb_Lb_L}^{\textsc{sm}}}{10^{-3}}\right)
\left(\frac{\delta g_{Z\mu_L\mu_L}/g_{Z\mu_L\mu_L}^{\textsc{sm}}}{5 \times 10^{-3}}\right)\,.
\end{equation}
This means that, in generic models without $P_{LR}$ protection,
sizable values for $C_{9,10}^\mu$ that could explain the $B$ anomalies are correlated
to deviations in the $Zb_L\overline b_L$, $Z\mu_L\overline \mu_L$, $Z\nu_\mu\overline \nu_\mu$ and $W\!\mu\nu_\mu$
couplings of the order of the present experimental bounds. This result strongly reduces the parameter space region
compatible with the $B$ anomalies, making the composite Higgs explanation a very predictive scenario.
We will see this mechanism at work in the explicit model we present in the next section.

\medskip

Since we are considering scenarios with a large lepton compositeness, we might wonder about possible large flavor
violating transitions in the lepton sector. Particularly dangerous are possible contributions to the $\mu \rightarrow e \gamma$
process, which imply a bound of tens of TeV on the mass scale of the resonances in anarchic composite Higgs
models~\cite{Panico:2015jxa}. To avoid these effects we need to assume that the rotations that diagonalizes the charged
lepton mass matrix are very close to the identity, so that flavor changing interactions with the vector resonances are not
generated. This can be obtained by imposing a $U(1)^3$ flavor symmetry in the lepton sector broken only by
the tiny effects due to the neutrino masses.

\section{An explicit model}

We now present an explicit model that can explain the $B$ anomalies. This scenario is analogous to the
usual RS set-up, the only difference being a modified background metric, which departs from conformality around the IR
brane. The details of the model have been discussed in ref.~\cite{Megias:2015ory}.
The metric has the form $ds^2 = e^{-2 A(y)} \eta_{\mu\nu} dx^\mu dx^\nu + dy^2$, where $\eta_{\mu\nu} = (-1,1,1,1)$
and $y$ is the coordinate along the extra dimension. The warp factor is determined by the dynamics of the scalar field
$\phi$ which stabilizes the size of the extra dimension. Its action has the form
\begin{equation}
S_\phi =  M^3\int d^4xdy\sqrt{-g}\left(R-\frac{1}{2}(\partial_M \phi)^2-V(\phi)\right)
-M^3 \sum_{\alpha}\int d^4x dy \sqrt{-g}\,2\mathcal V^\alpha(\phi)\delta(y-y_\alpha) \,,  \label{eq:model}
\end{equation}
where ${\mathcal V}^\alpha \; (\alpha=0,1)$ are the UV and IR brane potentials localized at $y_0 \equiv y(\phi_0)$
and $y_1 \equiv y(\phi_1)$ respectively, and $M$ is the 5D Planck scale.

The dynamics of $\phi$ can be described by a superpotential $W(\phi)$,
defined by~$V(\phi) \equiv \frac{1}{2} [W^\prime(\phi)]^2 - \frac{1}{3} W(\phi)^2$.~\cite{Gubser:2000nd} The background equations
then reduce to~$\dot{A}(y) = \frac{1}{6} W(\phi(y))$ and $\dot{\phi}(y) =  W^\prime(\phi)$, where
$\dot{X}\equiv dX(y)/dy$, and $Y^\prime\equiv dY(\phi)/d\phi$. 
The localization of the branes is governed by the effective potentials
$U_\alpha(\phi)\equiv\mathcal V_\alpha(\phi)-(-1)^\alpha W(\phi)$. The boundary conditions together with
the equations of motion lead to $U_\alpha(\phi)\big|_{y=y_\alpha} =U_\alpha^\prime(\phi)\big|_{y=y_\alpha}=0$.
In order to solve the Hierarchy Problem, the brane dynamics should fix $(\phi_0,\phi_1)$
to get $A(\phi_1) - A(\phi_0) \approx 35$.
We will fix $\phi_1=5$, while $\phi_0$ is used to fix the length of the extra-dimension.
In the following we assume the dynamics of $\phi$ to be characterized by the analytic superpotential
$W(\phi)=6k \big( 1+e^{a\phi} \big)$,
where $a$ is a real dimensionless parameter (which we set to $a=0.2$ for our numerical analysis), and $k$ is a mass parameter
related to the curvature along the fifth dimension.

We assume that a $5D$ gauge invariance is present, whose gauge group coincides with the SM one
$SU(3)_c \times SU(2)_L \times U(1)_Y$. 
In addition, we consider a Higgs field propagating in the bulk.
EWSB is triggered by an IR brane potential. The localization of the Higgs is controlled by the parameter $\alpha$
in the bulk mass term $M^2(\phi)=\alpha k \left[\alpha k-\frac{2}{3}W(\phi)  \right]$ and is connected
to the amount of tuning related to the Hierarchy Problem~\cite{Cabrer:2011fb}. Values $\alpha \gtrsim 3$ correspond to
a natural theory.

\begin{figure}
\centering
    \includegraphics[width=.375\textwidth]{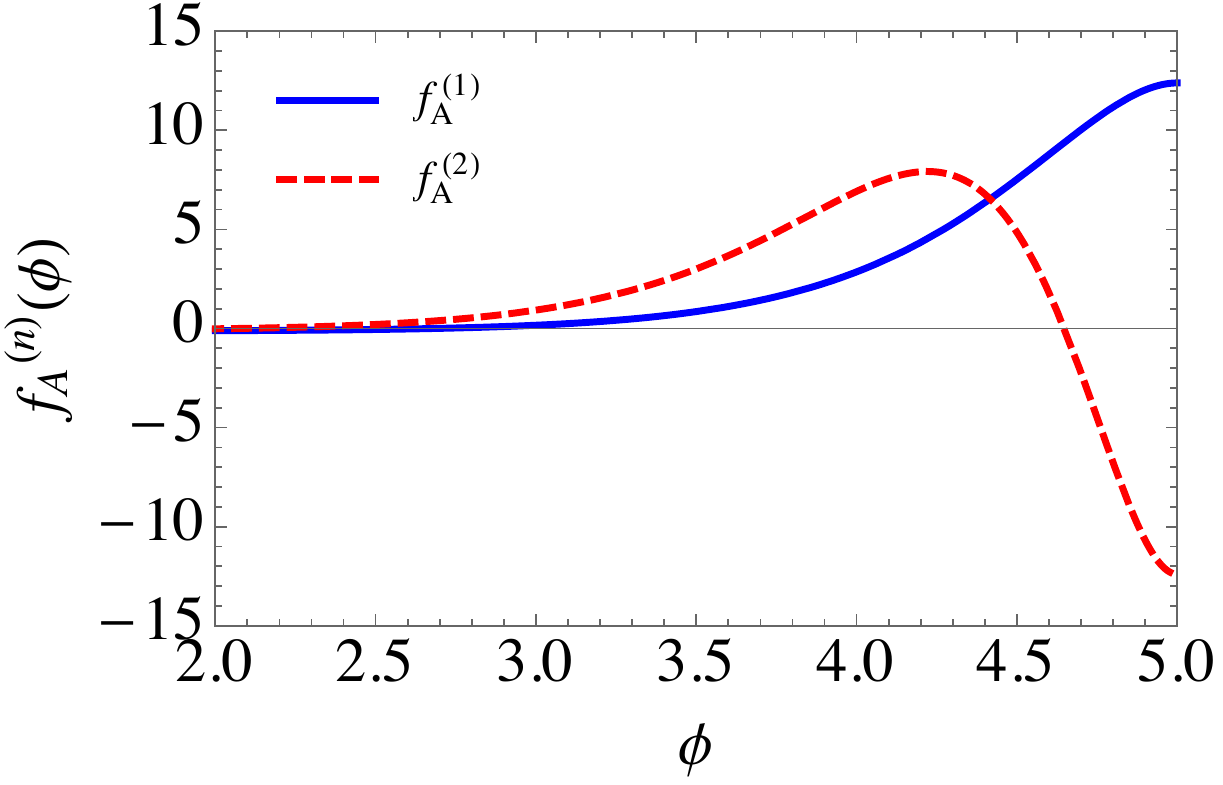}
    \hspace{2em}
     \includegraphics[width=.365\textwidth]{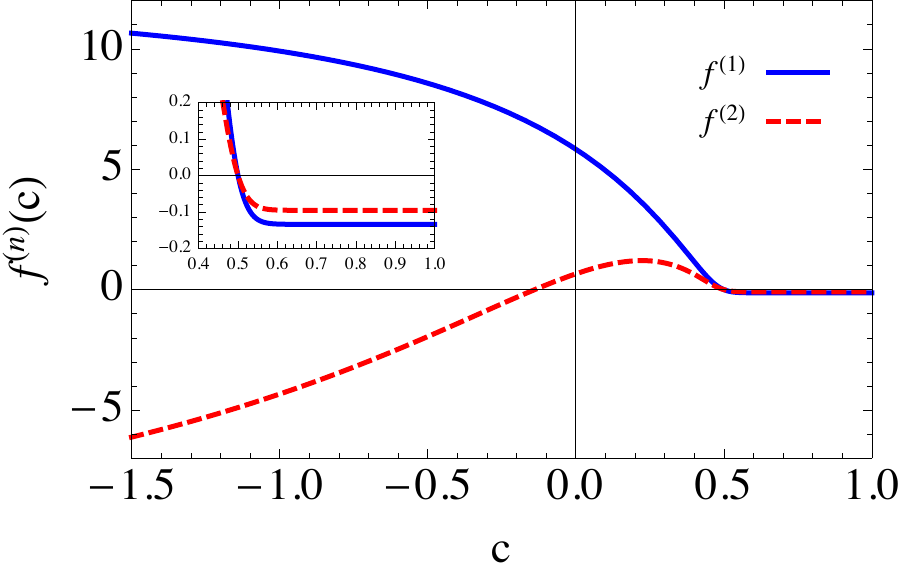}
\vspace{-0.3cm}
\caption{(Left) Profiles of the gauge boson KK modes $f_A^{(n)}$ for $n=1,2$ (solid blue and dashed red lines respectively). (Right) Coupling (normalized with respect to the 4D coupling $g$) of a fermion zero-mode with the $n$-th KK gauge field, $f^{(n)}(c)$, as a function of the fermion localization parameter $c$.}
\label{fig:KKcouplings}
\end{figure}

The gauge fields are decomposed in KK modes as $A_\mu(x,y)= \sum_n f^{(n)}_A(y) A^{n}_\mu(x)/\sqrt{y_1}$, where $f_A^{(n)}(y)$ satisfies Neumann boundary conditions and bulk equations $\big(m^{(n)}_A\big)^2f^{(n)}_A + \big( e^{-2A}\dot{f}^{(n)}_A \big)^{\dot{}} - M_A^2(y) f^{(n)}_A=0$, where $m_A^{(n)}$ denotes the mass of the $n$-th KK mode and $M_A(y)$ is the mass term induced by the vacuum expectation value of the Higgs~\cite{Cabrer:2011fb}.  We plot $f_A^{(n)}$ in fig.~\ref{fig:KKcouplings} (left).

The SM fermions are realized as chiral zero modes of 5D fermions. The localization of the different fermions is determined
by the 5D mass terms $M_{f_{L,R}}(y)=\mp c_{f_{L,R}} W(\phi)$~\cite{Cabrer:2011qb}.
The zero modes are localized near the UV (IR) brane for $c_{f_{L,R}}>1/2$ ($c_{f_{L,R}}<1/2$).
A value $c_{f_{L,R}} < 1/2$ thus corresponds to a sizable amount of compositeness for the corresponding fermions,
whereas $c_{f_{L,R}} > 1/2$ characterizes fermions that are almost elementary.
The coupling of the SM fermions with the massive KK modes of the gauge fields are universal and fully
determined by the localization of the fermions, i.e.~by the $c_{f_{L,R}}$ parameters.
The coupling with the $n$-th gauge KK mode, $X_\mu^n$,
can be written as~$g_{f_{L,R}}^{X^n}\,X_\mu^n \bar f_{L,R}\gamma^\mu f_{L,R} \equiv g f^{(n)}(c_{f_{L,R}})X_\mu^n\, \bar f_{L,R}\gamma^\mu f_{L,R}$, where $f_{L,R}$ are fermion zero-modes, $g$ is the SM gauge coupling and
$f^{(n)}(c_{f_{L,R}})$ encodes the overlap of the KK wave-function of the vector bosons with the zero mode fermion.
These functions are plotted in fig.~\ref{fig:KKcouplings} (right). Note that for almost elementary fields
the coupling becomes rather weak $\sim 0.1 g$.

\begin{figure}
\centering
      \raisebox{.5em}{\includegraphics[width=.39\textwidth]{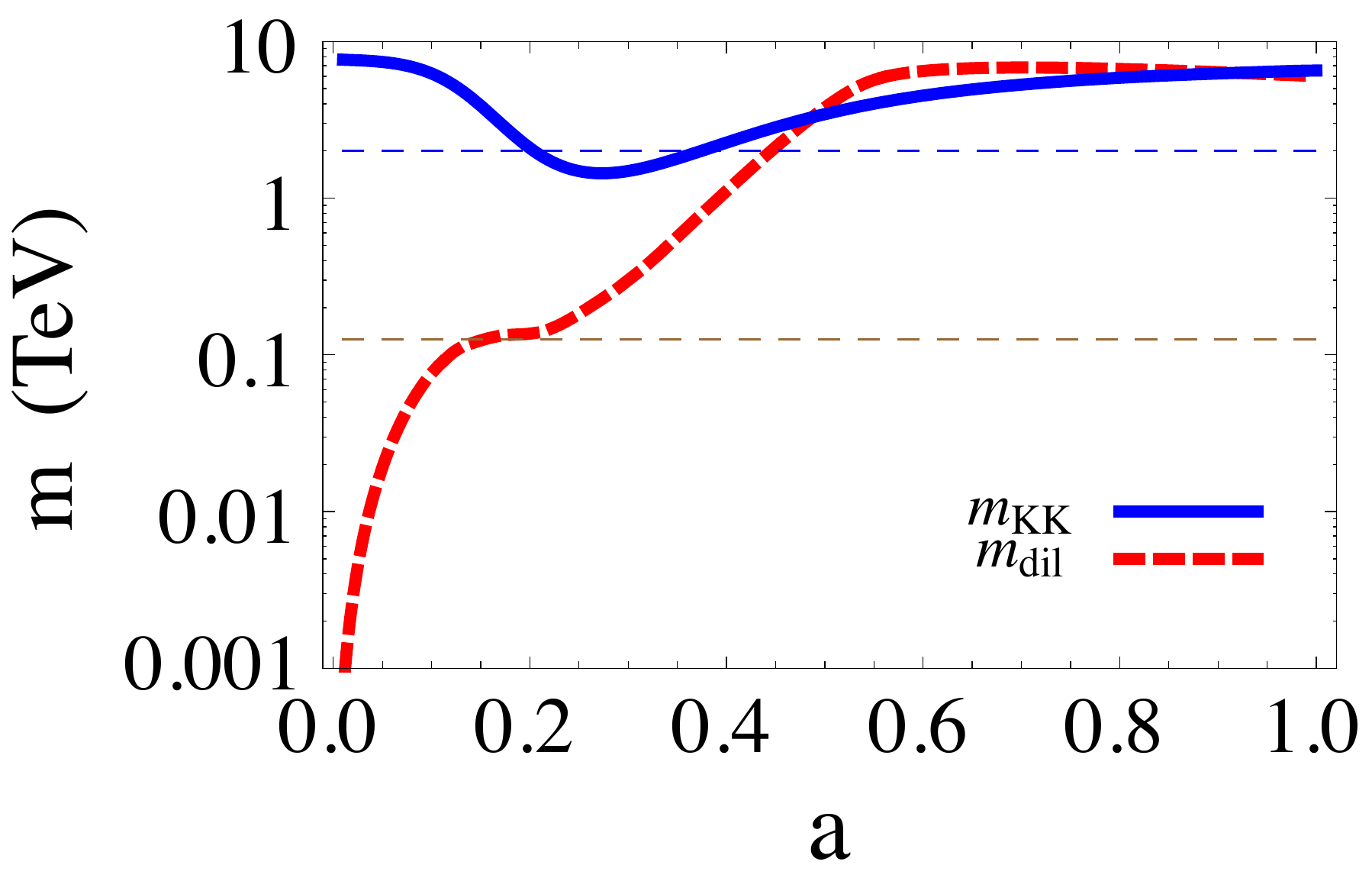}}
     \hfill
      \includegraphics[width=0.29\textwidth]{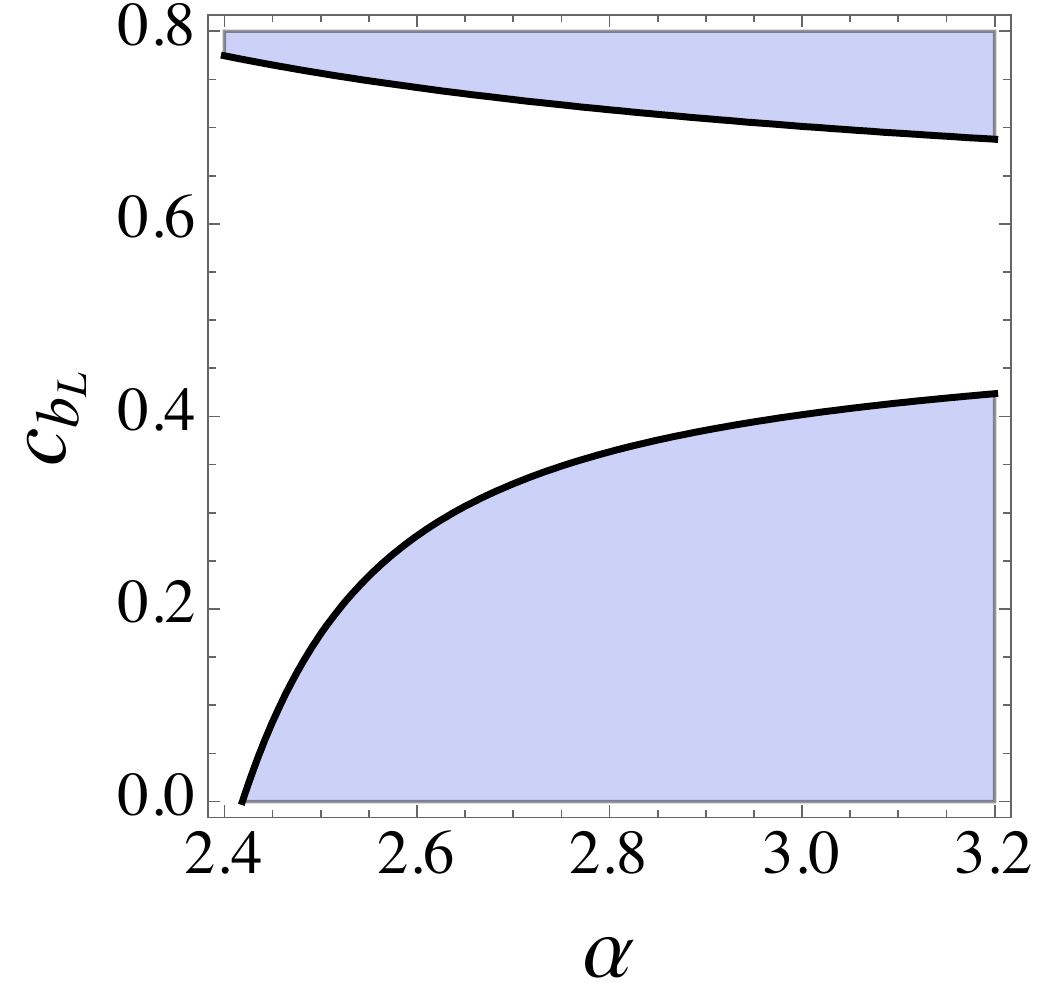} 
       \hfill
         \includegraphics[width=0.285\textwidth]{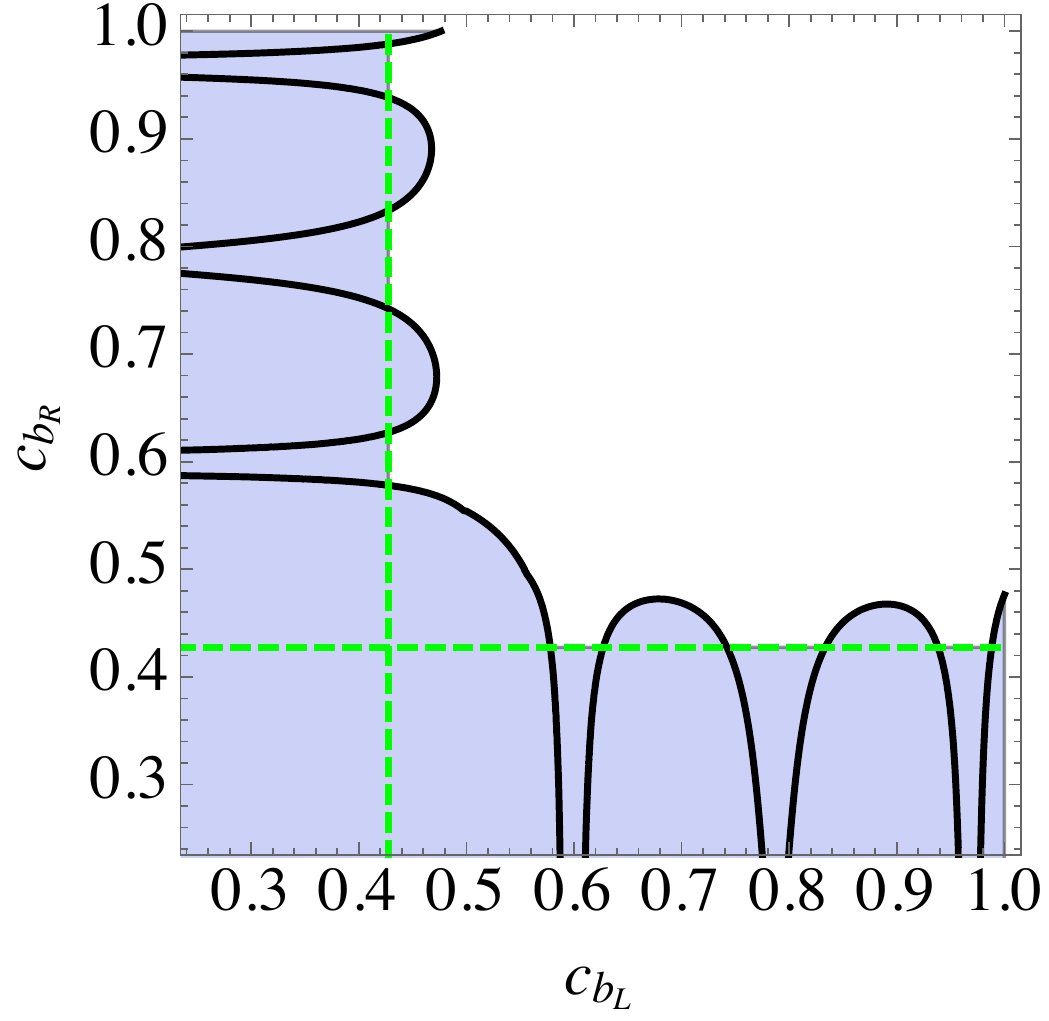}
\vspace{-0.3cm}
\caption{(Left) Lower bound on KK mass (solid line) as a function of the parameter $a$, computed from EW observables. The corresponding dilaton mass is in dashed line. The horizontal dashed lines correspond to $125$~GeV and $2$~TeV.
(Middle) Region in the plane $(\alpha,c_{b_L})$ allowed by experimental data on $\delta R_b^{exp} = 0.00053\pm 0.00066$ at the $3\sigma$ level. We have fixed $c_{b_R}=0.58$.
 (Right) Region in the plane $(c_{b_L},c_{b_R})$ that accommodates the bounds of Eq.~(\ref{eq:boundBs}). The dashed green lines
 represent $c_{b_{L,R}} = 0.43$. The allowed points correspond to the unshaded region.}
\label{fig:bounds}
\end{figure}

When comparing the model predictions with EW precision tests,
the most relevant bounds come from the oblique observables $S$ and $T$.
These constraints give a lower bound on the mass of the vector KK modes as well
as on the mass of the scalar mode (the dilaton).
The results are shown in fig.~\ref{fig:bounds}. We find that for $a \sim 0.3$ the KK-modes are allowed to have
a mass $m_{KK} = {\cal O}({\rm TeV})$. Interestingly in this region of the parameter space the model also predicts a light dilaton
with a mass $m_{\rm dil} \lesssim \mathcal O(500\ {\rm GeV})$. For the dilaton phenomenology see ref.~\cite{Megias:2015ory}.

\subsection{Reproducing the $B$ anomalies}
\label{sec:Banomaly}

As we discussed in the previous section, contributions to the ${\cal O}_{9,10}^{\mu}$ operators are generated by the
exchange of heavy vector resonances, in particular by the KK modes of the $Z$-boson and of the photon.
LFU can be broken by the different localization of the various lepton generations.
The leading flavor violating interactions with the vector KK modes have the form~\cite{Megias:2016bde}
\begin{equation}
\mathcal L_{EW}=\sum\!\rule[-.6em]{0pt}{.5em}_{X=Z,\gamma}\, \frac{X^n_\mu}{2c_W}\Big[
    V_{3i}^* V_{3j}\,\bar d_i\gamma^\mu \Big\{\left(g^{X^n}_{b_{L}}-\overline g^{X^n}_{L}  \right)P_L
+\left(g^{X^n}_{b_R}-\overline g^{X^n}_{R}  \right)P_R\Big\} d_j + {\rm h.c.} \Big]\,,\label{eq:lagrangianEW_od}
\end{equation}
where $c_W \equiv \cos \theta_W$, $P_{R,L} = (1 \pm \gamma_5)/2$, $V_{ij}$ are the CKM matrix elements,
and $\overline g^{X^n}_{L}$ are the couplings of $d_1$ and $d_2$ to the KK vectors.
The couplings in Eq.~(\ref{eq:lagrangianEW_od}) give rise to the contribution to the $C_{9,10}^{\mu}$ Wilson coefficients
\begin{equation}
\Delta C_9^{\mu} = -\Delta C_{10}^{\mu} = -\sum\!\rule[-.6em]{0pt}{.5em}_{X=Z,\gamma}\,\sum\!\rule[-.6em]{0pt}{.5em}_n\frac{\pi}{2\sqrt{2} G_F \alpha_{EM} c_W^2 M^2_n} g_{\mu_V}^{X_n}\left(g^{X_n}_{b_L}-g^{X_n}_{s_L}\right)\,.
\end{equation}
The largest contributions come from the exchange of the first KK excitations, $Z_\mu^1$ and $\gamma_\mu^1$.
The additional contributions are suppressed by the larger masses of the higher states, and lead to subleading corrections.

\subsection{Constraints}

The $Z$ boson couplings to SM fermions are modified by vector KK modes and fermion KK excitations.
After summing over the KK levels, the full result reads
\begin{equation}
\delta g_{b_{L,R}}= - g_{b_{L,R}}^{SM}m_Z^2\widehat \alpha_{b_{L,R}}\pm g {v^2}\widehat\beta_{b_{L,R}}/{2}\,,  \label{eq:delta_g}
\end{equation}
where $\widehat\alpha_{b_{L,R}}$ and $\widehat\beta_{b_{L,R}}$ are defined in ref.~\cite{Cabrer:2011qb}.
The main experimental constraints on the $Zb_L\overline b_L$ coupling come from the observables $R_b$, defined
as the ratio of the $Z\to b \overline b$ partial width to the inclusive hadronic width,
and $A_{FB}^b$, the forward-backward asymmetry of the bottom quark~\cite{Olive:2016xmw}.
We show in the middle panel of fig.~\ref{fig:bounds} how the bounds on $c_{b_L}$ vary as a function of the parameter $\alpha$,
which determines the amount of tuning in the Higgs sector. Values $\alpha \gtrsim 3$ correspond to a completely
natural theory, while $\alpha < 3$ corresponds to exponentially large tuning.
Analogously to $Zb_L \overline b_L$, the massive KK modes also induce modification on the muon couplings.
The result is obtained from Eq.~(\ref{eq:delta_g}) with obvious substitutions.
If we want to avoid fine tuning, the current bounds on the distortions of the muon coupling to the $Z$
implies~$c_{\mu_L}\gtrsim 0.4$. Similar bounds are obtained from the
distortion of the $Z\nu_\mu\overline \nu_\mu$ and $W\mu\nu_\mu$ couplings.

Another important set of constraints comes from $\Delta F = 2$ flavor-changing processes mediated by $4$-fermion interactions.
The main new physics contributions to these processes come from the exchange of gluon KK modes.
The current bounds on the $\Delta F = 2$ contact operators~\cite{Isidori:2015oea} can be translated into constraints on the quantities
\begin{equation}
\hspace{-0.6cm} \sum\!\rule[-.6em]{0pt}{.5em}_n {\big(g_{b_{L,R}}^{G^n}\big)^2}/{M_n^2[{\rm TeV}]}  \leq 0.14 \,, \qquad\quad \sum\!\rule[-.6em]{0pt}{.5em}_n {g_{b_L}^{G^n} g_{b_R}^{G^n}}/{M_n^2[{\rm TeV}]} \leq 3\times10^{-4} \,. \label{eq:boundBs}
\end{equation}
The first constraint leads to $c_{b_{L,R}} \ge 0.43$. The allowed configurations in the $(c_{b_L}, c_{b_R})$ plane are shown in
the right panel of fig.~\ref{fig:bounds}.

\section{Conclusions}

\begin{figure}
\centering
    \includegraphics[width=0.385\textwidth]{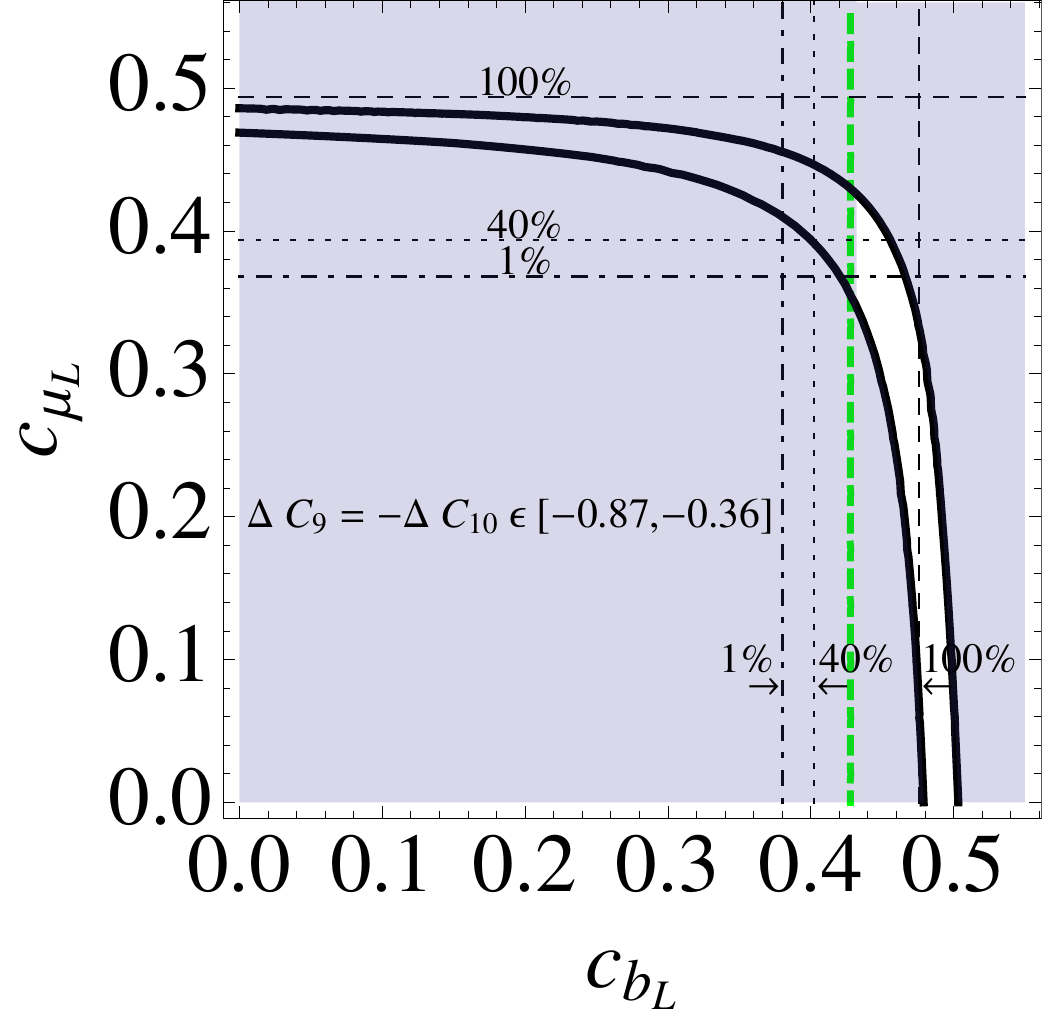} 
\vspace{-0.3cm}
\caption{Region in the plane $(c_{b_L},c_{\mu_L})$ that fits the $B$ anomalies. The region to the left of the vertical dashed green line is excluded by Eq.~(\ref{eq:boundBs}). The fine-tuning needed to pass the constraints on the modification of the $Z\mu_L\overline \mu_L$ ($Zb_L\overline b_L$) coupling is shown by the black dashed, dotted and dot-dashed horizontal (vertical) lines.
}
\label{fig:LHCbfinal}
\end{figure}

The results of our analysis are summarized in fig.~\ref{fig:LHCbfinal}, which shows the parameter space that allows
to fit the flavor anomalies. The horizontal and vertical black lines show the amount
of fine tuning in the Higgs sector needed to pass the EW constraints. A completely natural scenario corresponds to $100\%$,
whereas lines of $40\%$ and $1\%$ lead to a certain level of tuning.

We find that our extra-dimensional set-up can easily explain the anomalies in $B$-meson decays, as a direct consequence
of the LFU violation induced by a sizable compositeness for the left-handed bottom and muon components.
In agreement with the general estimates presented in section~\ref{sec:power-counting}, the interplay between the $B$-meson
data and the constraints from EW measurements (in particular the $Z$ couplings to the $b_L$ and $\mu_L$ and the $\Delta F=2$
transitions) singles out a preferred region of the parameter space in which all the bounds are satisfied with a small amount of tuning.
Incidentally, this region also predicts the presence of a light dilaton-like state, which could be detectable at hadron colliders.

To conclude we mention that also the anomalies found in $D$-meson decays by the BaBar, Belle and LHCb Collaborations
can be easily explained in our scenario by assuming a sizable compositeness for the $\tau_L$ field~\cite{Megias:2017ove}.

\section*{Acknowledgments}

G.~P. thanks the Organizers of the Moriond 2017 Conference for the kind invitation.
Work supported by MINECO Grant CICYT-FEDER-FPA2014-55613-P, FPA2015-64041-C2-1-P, Severo Ochoa Excellence Program Grant SO-2012-0234, and by Generalitat de Catalunya Grant 2014 SGR 1450 and by the Basque Government under Grant IT979-16.
The research of E.M. is supported by the European Union (FP7-PEOPLE-2013-IEF) project PIEF-GA-2013-623006, and by the
Universidad del Pa\'{\i}s Vasco UPV/EHU, Bilbao, Spain, as a Visiting Professor.

\end{document}